\documentclass[twocolumn,           
               showpacs,            
               preprintnumbers,     
               aps,                 
               prd,          	    
               letterpaper,             
               superscriptaddress,      
               nofootinbib,         
               tightenlines,        
               floats,floatfix      
               ]{revtex4-1}

\usepackage{graphicx}
\usepackage{dcolumn}
\usepackage{bm}
\usepackage{amsmath}
\usepackage{epstopdf}

\begin{document}

\title{New perturbative method for analytical solutions in
  single-field models of inflation}

\author{L. Arturo Ure\~na-L\'opez}%
 \email{lurena@ugto.mx}
\affiliation{%
Departamento de F\'isica, DCI, Campus Le\'on, Universidad de
Guanajuato, 37150, Le\'on, Guanajuato, M\'exico.}

\date{\today}

\begin{abstract}
We propose a new parametrization of the background equations of
motion corresponding to (canonical) single-field models of inflation,
which allows a better understanding of the general properties of the
solutions and of the corresponding predictions in the inflationary
observables. Based on the tools of dynamical systems, the method
suggests that inflation comes in two flavors: power-law and de
Sitter. Power-law inflation seems to occur for a restricted type of
potentials, whereas de Sitter inflation has a much broader
applicability. We also show a general perturbative method, by means of
series expansion, to solve the new equations of motion around the
critical point of the de Sitter type, and how the method can be used
for arbitrary models of de Sitter inflation. It is then argued that
for the latter there are two general classes of inflationary solutions,
given in terms of the behavior of the tensor-to-scalar ratio as a
function of the number $N$ of $e$-folds before the end of inflation:
$r \sim N^{-1}$ (Class I), or $r \sim N^{-2}$ (Class II). We give some
examples of the two classes in terms of known scalar field potentials,
and compare their general predictions with constraints obtained from
observations.
\end{abstract}

\maketitle

\section{Introduction \label{sec:introduction-}}
Inflation remains one of the cornerstones of modern
cosmology, and it still has a prominent place within the big bang
model for the evolution of the Universe. According to recent accounts
from the cosmic microwave background (CMB) measurements by the Planck
Collaboration, single-field models of inflation are still among the
best candidates to explain the features of the primordial power
spectrum of density perturbations\cite{Ade:2015lrj}.

Large efforts have been dedicated to solving the equations of motion of
inflationary scalar fields under different techniques, and one of the most
successful of them is the well-known slow-roll
approximation\cite{Liddle:2000cg,*Mukhanov:2005sc}. The two basic
assumptions for slow roll are that the energy density is dominated by
the scalar potential, and that the scalar field itself is evolving
slowly enough to guarantee an extended period of accelerated
expansion. The equations of motion for the background can then be
written in a closed form that allows general solutions, up to
quadratures, of the scalar field as a function of the number of
$e$-foldings of expansion for any scalar potential. The background
solution is then used to calculate different features of the linear
field perturbations that can be compared with actual
observations. Although the general solution of the field perturbations
is the same for any scalar field model of inflation, the final output
depends on the specific background evolution, and it is because of
this that observations may potentially distinguish among different
inflationary models\cite{Malik:2008im}.

It is then necessary to have a good understanding of the background
evolution for different models if one is to find the one preferred by
the observations. Even though good numerical resources are now at hand\cite{Mortonson:2010er}, the thorough studies
compiled in Enciclopaedia
Inflationaris\cite{Martin:2014vha,*Martin:2015dha}, and in
Enciclopaedia Curvatonis\cite{Vennin:2015vfa,*Vennin:2015egh}, show
that better formulas of the inflationary solutions are still needed
and much appreciated, as they still offer a clearer connection with
physical parameters and observable quantities that may not be easily
obtained from numerical simulations.

The aim of this paper is to present a new and appropriate set of
equations that can allow a proper understanding of the subtleties of
inflationary background dynamics, so that useful and appropriate
solutions can be obtained for analytical and numerical purposes. This
will be achieved through the use of the tools of dynamical
systems (for some pedagogical presentations,
see Ref.~\cite{UrenaLopez:2006ay,*Faraoni:2012bf,*Garcia-Salcedo:2015ora}),
together with appropriate changes of variables that are more than
suitable to exploit the intrinsic symmetries of the equations of
motion. Even though we will be mainly concerned with single-field
models of inflation, the techniques that will be developed here have
been recently extended to different cosmological settings with scalar
fields\cite{Urena-Lopez:2015gur}.

A brief description of the paper is as follows. In
Sec.~\ref{sec:scal-field-dynam} we explain the conversion of the
equations of motion for a scalar field-dominated Universe into an
autonomous dynamical system, the latter of which is written in terms of a
kinetic variable and two potential variables. Taking into account the
symmetries of the system of equations, we propose a change to polar
coordinates that helps to convert them into a more manageable
two-dimensional dynamical system that can be solved perturbatively by
means of a series expansion. It is then shown that there is one type
of critical point of the dynamical system that corresponds to a de
Sitter phase, and of which the stability properties can be easily described
in general terms for any kind of potential. The perturbative solutions
will result in a better description of the inflationary dynamics phase
than the usual slow-roll approximation, and then useful analytical
formulas can be constructed to give a more precise determination of
different inflationary quantities.

In Sec.~\ref{sec:general-solutions-} we show the general procedure to
generate series solution at any given order, even though we
will focus our attention to the third order, which will be enough for
most cases. We will also present the general procedure to apply the
perturbative method to arbitrary scalar field models. In
Sec.~\ref{sec:gener-class-solut}, we will explain that there seems to
be two general classes of  models, which depend upon the values of one
of the constant coefficients of the series expansion in the
solutions. We then give some particular examples of each case for
illustration purposes, and explain their general predictions for the
inflationary observables. The second and most general class is shown
for completeness, although a particular example remains to be found
for it. In Sec.~\ref{sec:comp-with-observ} we make a general
comparison of the two classes of solutions with the observational
constraints imposed by the Planck Collaboration
results\cite{Ade:2015lrj} on both the spectral index of density
perturbations and the tensor-to-scalar ratio. Finally, in
Sec.~\ref{sec:conclusions} we give a summary of the main results.

\section{Scalar field dynamics \label{sec:scal-field-dynam}}
Let us consider the background equations of motion for a
spatially flat Universe dominated by a scalar field $\phi$ endowed
with a potential $V(\phi)$\cite{Liddle:2000cg,*Mukhanov:2005sc}:
\begin{subequations}
  \label{eq:1}
  \begin{eqnarray}
    H^2 &=& \frac{\kappa^2}{3} \left[ \frac{1}{2} \dot{\phi}^2 +
      V(\phi) \right] \, ,   \label{eq:1a} \\
    \dot{H} &=& - \frac{\kappa^2}{2} \dot{\phi}^2 \, , \label{eq:1b}
    \\
    \ddot{\phi} &=& -3 H \dot{\phi} - \partial_\phi V(\phi) \,
    , \label{eq:1c}
  \end{eqnarray}
\end{subequations}
where $\kappa^2 = 8\pi G$, a dot denotes derivative with respect to
cosmic time, and the background spacetime is described by the
Friedmann-Lemaitre-Robertson-Walker metric, with $H$ its Hubble
parameter. Following the seminal paper\cite{Copeland:1997et}, the full
system~\eqref{eq:1} can be written as a dynamical system if we define
the variables
\begin{subequations}
\label{eq:2}
  \begin{eqnarray}
    x \equiv \frac{\kappa \dot{\phi}}{\sqrt{6} H} \, , \quad y
    \equiv \frac{\kappa V^{1/2}}{\sqrt{3} H} \, , \label{eq:2a} \\
    y_I \equiv - \frac{2 \kappa}{\sqrt{3}}
    \frac{6^{I/2}}{\kappa^I} \frac{\partial^I_\phi V^{1/2}}{H} \,
    , \label{eq:2b}
  \end{eqnarray}
\end{subequations}
which is appropriate as long as the scalar potential $V(\phi)$ is
positive definite. Here the label $I \geq 1$ denotes the derivative
order in Eq.~\eqref{eq:2b}. As a result, the Klein-Gordon equation can
be written as an \emph{extended} system of first-order differential
equations:
\begin{subequations}
\label{eq:3}
  \begin{eqnarray}
    x^\prime &=& 3(x^2 -1) x + \frac{1}{2} y y_1 \, , \label{eq:3a} \\
    y^\prime &=& 3 x^2 y - \frac{1}{2} x y_1 \, , \label{eq:3b} \\
    y^\prime_I &=& 3 x^2 y_I + x y_{I+1} \, , \quad I \geq 1 \,
                   , \label{eq:3c}
  \end{eqnarray}
\end{subequations}
where a prime denotes derivative with respect to the number of
$e$-foldings $N$, and $y_I$ in Eq.~\eqref{eq:3c} is defined as in
Eq.~\eqref{eq:2b}. In writing Eqs.~\eqref{eq:3} we have made use of
the acceleration equation~\eqref{eq:1b} in the form $\dot{H}/H^2 = -3
x^2$.

Notice that the kinetic and potential variables move in the finite
ranges $x= [-1,1]$ and $y= [0,1]$\footnote{The negative branch of $y$
  is excluded because we are only interested in solutions for an
  expanding Universe with $H>0$.}, but the other potential variables
$y_I$ have in principle an open range of variation. The ranges for $x$
and $y$ are inferred from the extreme values they can take to saturate
the Friedmann constraint~\eqref{eq:1a}. The latter actually reads:
$x^2+y^2=1$, and then we can have $x=\pm 1$ for a kinetic dominated
expansion, or $y= 1$ for a potential-dominated one.

We now apply a \emph{polar} change of variables to the kinetic and
potential variables in the form: $x =\sin(\theta/2)$ and $y
=\cos(\theta/2)$, which automatically guarantees the accomplishment
of the Friedmann constraint. This type of transformation was first
proposed in Ref.~\cite{Reyes:2010zzb}, and recently applied to scalar field
models of dark matter in\cite{Urena-Lopez:2015gur}; but see also Refs.~\cite{Linder:2010aa,Rendall:2006cq,*Alho:2014fha} for other similar
definitions. Thus, after straightforward manipulations
the first three equations in~\eqref{eq:3} finally become
\begin{subequations}
\label{eq:4}
  \begin{eqnarray}
  \theta^\prime &=& -3 \sin \theta + y_1 \, , \label{eq:4a} \\
  y^\prime_1 &=& \frac{3}{2}\left( 1 - \cos \theta \right) y_1 +
  \sin(\theta/2) y_2 \, . \label{eq:4b}
\end{eqnarray}
\end{subequations}
Equation~\eqref{eq:4a} results from the combination of Eqs.~\eqref{eq:3a}
and~\eqref{eq:3b}, whereas Eq.~\eqref{eq:4b} is just a direct
rewriting of Eq.~\eqref{eq:3c}. The rest of Eqs.~\eqref{eq:3} with $I
\geq 2$ remain the same. The equation of state (EoS) of the scalar
field is given by $w_\phi = (x^2 - y^2)/(x^2 + y^2) = - \cos \theta$.\footnote{Interestingly enough, a version of Eq.~\eqref{eq:4a}
  written in terms of the scalar EoS $w_\phi$ can be found
  in\cite{Linder:2010aa,Caldwell:2005tm} for the study and general
  classification of dark energy models with quintessence scalar
  fields.} The angular variable $\theta$ then gives direct
information about the EoS, but also about the ratio of the kinetic to
potential energies via the trigonometric identity $x/y =
\tan(\theta/2)$. One important point to notice is that
Eq.~\eqref{eq:4a} has the same form for all models of (canonical)
inflation, and it is only Eq.~\eqref{eq:4b} that will change for
different potentials through the definitions of $y_1$ and $y_2$ given in
Eq.~\eqref{eq:2b}.

\subsection{Perturbative solutions and inflationary quantities \label{sec:pert-solut-infl}}
In any given inflationary setting, we expect the scalar field to
change slowly from a potential to a kinetic-dominated phase, which in
terms of the EoS means $w_\phi:-1 \to 1$. Actually, it is only
required that $w_\phi:-1 \to -1/3$ to span an accelerating phase,
which in terms of the angular variable translates into $\theta:0 \to
\theta_{end}$, where $\theta_{end} = \arccos(1/3) =
2\arcsin(1/\sqrt{3})$. Notice that $\theta_{end} \simeq 1.2309 \ldots$
is then a fixed and the same number for all models of inflation.

The above discussion suggests that $\theta$ remains reasonably small
during an inflationary phase, so that we can try a series solution of
Eqs.~\eqref{eq:4} with the ansatz for the potential
variables
\begin{equation}
  y_I = \sum_{j=0} k_{Ij} \theta^j \, , \quad k_{Ij} =
  \mathrm{const}. \, , \label{eq:5}
\end{equation}
where the only exception is $k_{10}=0$. The justification for this
ansatz is given in Sec.~\ref{sec:general-solutions-} below, and for
now we would like to emphasize that, in principle, the expansion of
$y_1$ is all that is needed to find a full solution of the scalar field
equations~\eqref{eq:4}, and in turn also of the full background
one~\eqref{eq:1}. Indeed, once we have the values of $k_{1j}$ at
hand we can make an expansion of Eq.~\eqref{eq:4a} in the form:
\begin{equation}
  \theta^\prime = (k_{11} -3)\theta + k_{12} \theta^2 + (k_{13} + 1/2)
  \theta^3 + \ldots \, . \label{eq:6}
\end{equation}
Equation~\eqref{eq:6} can then be solved at any order to provide a solution:
$\theta_N = \theta_N (k_{1j}, \theta_{end},N)$, where subscript $N$
here denotes the number of $e$-foldings before the end of inflation.

To better understand this proposal, let us review the values of
inflationary quantities in terms of the new dynamical variables
$\theta$ and $y_1$. For that, we choose the so-called Hubble slow-roll
(HSR) variables\cite{Liddle:1994dx}, which appear to be well suited
for the dynamical system approach adopted in this paper. Explicitly,
we have that
\begin{subequations}
\label{eq:7}
  \begin{eqnarray}
  \epsilon_H &\equiv& 3 \frac{\dot{\phi}^2/2}{\dot{\phi}^2/2 + V} = 3
  \sin^2(\theta/2) \, , \\ 
  \eta_H &\equiv& - 3 \frac{\ddot{\phi}}{3H \dot{\phi}} = 3 -
  \frac{y_1}{2} \cot(\theta/2) \, .
\end{eqnarray}
\end{subequations}
Various inflationary observables can be written in terms of the
HSR parameters $\epsilon_H$ and $\eta_H$. For our purposes, it
suffices to consider the spectral index $n_S$, and the
tensor-to-scalar ratio $r$, which are given at first order in the HSR
variables as
\begin{subequations}
\label{eq:8}
  \begin{eqnarray}
    1-n_S &=& 4 \epsilon_H - 2 \eta_H = 12 \sin^2(\theta/2) - 6 + y_1
    \cot(\theta/2) \, , \\ 
    r &=& 16 \epsilon_H = 48 \sin^2(\theta/2) \, .
  \end{eqnarray}
\end{subequations}
Taking into account the expansion~\eqref{eq:5}, Eqs.~\eqref{eq:8} can
be written alternatively as
\begin{subequations}
  \label{eq:9}
  \begin{eqnarray}
    1 - n_S &\simeq& 2(k_{11} -3) + 2 k_{12} \theta_N + (2k_{13} -
    k_{11}/6 + 3) \theta^2_N  \label{eq:9a} \\
    r &\simeq& 12 \theta^2_N \, ,  \label{eq:9b}
  \end{eqnarray}
\end{subequations}
where we have considered an expansion up to the second order in
$\theta_N$ for $n_S$, which is also the lowest order in the expansion
of $r$.

Equations~\eqref{eq:9} represent a parametric curve on the inflationary
plane $(n_S,r)$, the exact form of which depends upon the particular
features of the model expressed only through the expansion coefficients
$k_{1j}$ of the potential variable $y_1$.

\subsection{Critical points and their stability \label{sec:crit-points-stab}}
As for any dynamical system, we investigate here the critical points
$(\theta_c,y_{1c},y_{2c})$ for which $\theta^\prime = 0 = y^\prime_1$ in Eqs.~\eqref{eq:4}. These are found from the conditions
\begin{subequations}
  \label{eq:10}
\begin{eqnarray}
  y_{1c}- 3 \sin \theta_c &=& 0 \, ,  \label{eq:10a} \\ 
  \left[ 3 \cos(\theta_c/2) y_{1c} + y_{2c} \right] \sin (\theta_c/2)
  &=& 0 \, .  \label{eq:10b}
\end{eqnarray}
\end{subequations}

The first critical point corresponds to $\theta_c \neq 0$, and in
this respect corresponds to power-law inflation with $a \sim
t^m$ and $m > 1$\cite{Liddle:2000cg,Lucchin:1984yf,*Ringstrom:2009zz}. In fact, the
latter means that $\dot{H}/H^2 = -3 x^2_c = - 3 \sin^2(\theta_c/2)=
-1/m = \mathrm{const}$, and then the inflationary expansion never
ends. The solution of $\theta_c$ must be found from
Eq.~\eqref{eq:10b}:
\begin{subequations}
\begin{equation}
  3 \cos(\theta_c/2) y_{1c} + y_{2c} =0 \, .
\end{equation}
From the expansions~\eqref{eq:5}, we find that if this critical point
also belongs to the inflationary solution, then the expansion
coefficients of $y_1$ and $y_2$ must be related through $k_{2j}=-3
\cos(\theta_c/2) k_{1j}$, in which we must take into account that
possibly $k_{10}\neq 0$. That is, the potential variables
themselves must be related one to one another in a very particular way,
more precisely $y_2= -3 \cos(\theta_c/2) y_1$. 

Indeed, from Eqs.~\eqref{eq:2b}, the only case in which this happens
is for the type of potentials:
\begin{equation}
  \label{eq:22}
  V(\phi) = M^4 \left(e^{-\lambda \kappa \phi} + \hat{M}^2 \right)^2
  \, ,
\end{equation}
\end{subequations}
where $\lambda = 3 \cos(\theta_c/2)/\sqrt{6}$, $M$ is the energy
scale of the potential, and $\hat{M}$ is a dimensionless
constant. Notice that the standard exponential potential of
power-law inflation is recovered if
$\hat{M}=0$\cite{Liddle:2000cg}. Because of this, we do not expect
inflation to generically be power law, but rather to be of the de
Sitter type (see below), as the latter does not impose any \emph{a
  priori} relationship between the derivatives of the scalar field
potential.

The second critical point corresponds to $\theta_c=0$, under which
$y_{1c}=0$, but the value $y_{2c}$ is left undetermined. The
corresponding value of the EoS is $w_\phi=-1$, which means that the
scalar field potential dominates the energy budget, and then the
critical point represents a de Sitter point. Notice also that this
directly indicates, through the expansion in Eq.~\eqref{eq:5} for
$y_1$, that $k_{10}=0$, if the inflationary solution we are looking
for is to contain the de Sitter critical point. 

To investigate the stability of the critical points we perform small
perturbations about the critical values in the form $\theta =
\theta_c + \delta \theta$, $y_1 = y_{1c} + \delta y_1$, and $y_2 =
y_{2c} + \delta y_2$. The linear version of Eqs.~\eqref{eq:4} around
the de Sitter point is then:
\begin{equation}
  \left( 
    \begin{array}{c}
      \delta \theta \\
      \delta y_1
    \end{array}
  \right)^\prime = \left(
    \begin{array}{cc}
      -3 & 1 \\
      k_{20}/2 & 0
    \end{array}
  \right) \left( 
    \begin{array}{c}
      \delta \theta \\
      \delta y_1
    \end{array}
  \right) \, , \label{eq:11}
\end{equation}
where we have considered that $y_{2c}= k_{20}$ is the only surviving
term in the expansion of $y_2$ as $\theta \to 0$; see
Eqs.~\eqref{eq:2b}. The perturbation $\delta y_2$ does not appear
explicitly in the linear equations~\eqref{eq:10}, and then the
stability of the critical point can be determined on general terms for
all potentials.

If we look for a solution of Eqs.~\eqref{eq:11} in the form $\sim
e^{\omega N}$, the eigenvalues of $\omega$ are
\begin{equation}
  \omega_\pm = \frac{3}{2} \left( \pm \sqrt{1+ \frac{2}{9} k_{20}}
    - 1 \right) \, . \label{eq:12}
\end{equation}
This means that $\omega_-$ represents the decaying mode of the linear
solution, as its real part will always be negative irrespective of the
value and sign of $k_{20}$. The same does not apply for $\omega_+$,
the real part of which can be either positive or negative, and this depends
upon the overall sign of $k_{20}$. If the critical point is going to
represent an unstable stage of inflation, then in general terms we
must expect $k_{20} \geq 0$ so that $\omega_+ > 0$; in such a case, the
critical point will be a saddle. 

The full linear solution reads
\begin{equation}
  \left( \begin{array}{c}
    \theta \\
    y
  \end{array} \right) = C_1 \left(
  \begin{array}{c}
    1 \\
    -\omega_-
  \end{array} \right) e^{\omega_+ N} + C_2 \left(
  \begin{array}{c}
        1 \\
    -\omega_+
  \end{array} \right) e^{\omega_- N} \, , \label{eq:24}
\end{equation}
where $C_1$ and $C_2$ are integration constants that depend upon the initial
conditions. Equation~\eqref{eq:24} indicates that the surviving mode at
late times is the growing one that corresponds to unstable eigenvalue
$\omega_+$, and then at linear order the dynamical variables must be
related through the relation $y_1 \simeq - \omega_- \theta$. The
latter expression, in combination with the expansion of $y_1$ in
Eq.~\eqref{eq:5}, readily indicates that $k_{11} = - \omega_-$, and
then we see from Eq.~\eqref{eq:12} that $k_{11} \geq 3$, just because
of the aforementioned condition $k_{20} \geq 0$ for the instability of
the critical point.

To close this section, it must be stressed that the linear solution
$y_1 = k_{11} \theta$ is the solution of a very particular trajectory
that departs from the critical point. In the language of dynamical
systems, it is called an heteroclinic line, and it is this type of
line that plays the role of attractor solutions. As shown
in Ref.~\cite{UrenaLopez:2011ur} for the case of quintessence fields, in
order to have a complete description of the scalar field dynamics, it
is just enough to find the heteroclinic solutions that depart from the
critical points of interest. Any other solution with different initial
conditions, if given enough time, will join asymptotically the
(heteroclinic) attractor solutions. In the present case, we will find
a solution to the heteroclinic line that departs from the de Sitter
critical point, and this solution will be identified with the
inflationary solution of the scalar field equations of
motion~\eqref{eq:1}.

\begin{widetext}
\subsection{Interpretation of the series expansion \label{sec:interpr-seri-expans}}
The series expansions~\eqref{eq:5} have a well-defined meaning: they
are the Taylor expansions of the evolution of the potential variables
$y_I$ along the attractor trajectory that departs from the de Sitter
critical point $(\theta_c=0,y_{1c}=0)$. That is, they only represent a
particular parametrization of the time evolution of the potential
variables when normalized by the Hubble parameter [see
Eqs.~\eqref{eq:2b}]. 

There is, although, a second interpretation of the expansion coefficients
$k_{Ij}$. With the help of the first potential variable $y$, we can
write
\begin{subequations}
\label{eq:35}
  \begin{eqnarray}
    2 \frac{\partial_\phi V^{1/2}}{V^{1/2}} &=& \partial_\phi \left[ \ln V
    \right] = - \frac{\kappa}{\sqrt{6}} \frac{y_1}{y} = -
                                                \frac{\kappa}{\sqrt{6}}
                                                \left[ k_{11} \theta +
    k_{12} \theta^2 + \left( \frac{k_{11}}{8}+k_{13} \right) \theta^3 +
    \ldots \right] \, , \label{eq:28} \\
    \frac{\partial^I_\phi V^{1/2}}{V^{1/2}} &=&  - \frac{\kappa^I}{2
    \sqrt{6^I}} \frac{y_I}{y} = - \frac{\kappa^I}{2 \sqrt{6^I}} \left[
    k_{I0} + k_{I1} \theta + \left( \frac{k_{I0}}{8} + k_{I2} \right)
    \theta^2 + \left( \frac{k_{I1}}{8} + k_{I3} \right) \theta^3 +
    \ldots \right] \, , \quad I \geq 2 \, . \label{eq:33}
  \end{eqnarray}
\end{subequations}
From Eqs.~\eqref{eq:35} we see that the Hubble parameter can be left
out of the calculations, and the coefficients $k_{Ij}$ then have a
direct connection with the potential and its derivatives. In
particular, the coefficients $k_{1j}$ arise from the Taylor expansion
of the logarithmic derivative of the potential around the de Sitter
critical point. In principle, Eq.~\eqref{eq:28} can also be inverted
to find the corresponding parametrization of the scalar field itself
in terms of the angular variable, so that we can write
$\phi=\phi(\theta)$.

All other equations~\eqref{eq:33} also represent the parametrization
of the derivatives of higher order of the potential, and in principle
we would have to find all of them in order to have a complete picture
of the scalar field dynamics. However, because the field derivatives
$\partial^I_\phi V$ must be somehow related to one another, the
coefficients $k_{Ij}$ are not all independent, and the relations among
them, as we will see below, will depend on the particular form of the
potential.

\section{General inflationary solutions \label{sec:general-solutions-}}
Here we will explain how to solve Eqs.~\eqref{eq:4} for general
inflationary trajectories that depart from the de Sitter point at
$\theta = 0 = y_1$, and will also show some generic predictions of the
method. To have a good approximation to the full solution, it
is necessary to keep in Eq.~\eqref{eq:6} the expansions up to the
third order at least, and then we must impose that $k_{13} \neq
0$.\footnote{In providing general solutions, we should be aware of the
  limitations of the perturbative expansions~\eqref{eq:5} of the
  potential variables $y_I$. Since the end value of the angular
  variable $\theta_{end}$ is of order of unity, an expansion up to the
  order of $\theta^3$ in Eq.~\eqref{eq:6} may not in all cases be
  sufficient to guarantee enough accuracy in the solutions. However,
  we will show in Sec.~\ref{sec:gener-class-solut} below that the
  analytical solutions~\eqref{eq:27} are appropriate for our purposes
  (see also Appendix~\ref{sec:notes-accur-seri}).}

The general solution of Eq.~\eqref{eq:6} up to the second (2nd) and
third (3rd) order, respectively, are
\begin{subequations}
\label{eq:27}
\begin{eqnarray}
   \theta^{(2nd)}_N &=& \theta_{end} \theta_1 \left[ (\theta_{end} +
     \theta_1) e^{(k_{11}-3)N} - \theta_{end} \right]^{-1} \,
                        . \label{eq:27a} \\
  (k_{11}-3) N^{(3rd)} &=& \ln \left( \frac{\theta_{end}}{\theta_N}
  \right) + \frac{\theta_-}{\theta_+-\theta_-}
  \ln \left( \frac{\theta_{end} - \theta_+}{\theta_N - \theta_+}
  \right) - \frac{\theta_+}{\theta_+-\theta_-}
  \ln \left( \frac{\theta_{end} - \theta_-}{\theta_N - \theta_-}
  \right) \, , \label{eq:27b}
\end{eqnarray}
where
\begin{equation}
  \label{eq:27c}
  \theta_\pm = \frac{- k_{12} \pm \sqrt{k^2_{12} - 2(k_{11}-3)(2
      k_{13} +1)}}{2 k_{13}+1} \, , \quad \theta_1 =
  \frac{k_{11}-3}{k_{12}} \, .
\end{equation}
\end{subequations}

Strictly speaking, Eq.~\eqref{eq:27} is just a solution with no
information about the inflationary nature of the calculations we are
interested in. To make a connection with the original
equations of motion~\eqref{eq:1}, we must find the correct coefficients
$(k_{11},k_{12},k_{13})$.  The first step key to doing so is to write
Eq.~\eqref{eq:4b} in the form:
\begin{equation}
  y^\prime_1 = \frac{dy_1}{d\theta} \theta^\prime =
  \frac{dy_1}{d\theta} (-3\sin \theta + y_1) = \frac{3}{2}\left( 1 -
    \cos \theta \right) y_1 + \sin(\theta/2) y_2 \, , \label{eq:13}
\end{equation}
After expanding the functions $y_1$ and $y_2$ as in Eqs.~\eqref{eq:5}, and
also the sine and cosine functions, in powers of $\theta$, the
resultant polynomials on both sides of Eq.~\eqref{eq:13} must have the
same coefficients, and from this we find the following relationships
for the coefficients $k_{1j}$ and $k_{2j}$ up to the third order:
\begin{subequations}
  \label{eq:14}
  \begin{eqnarray}
    k_{11} (k_{11} - 3) &=& \frac{k_{20}}{2} \, , \label{eq:14a} \\
    k_{12} (k_{11} - 2) &=& \frac{k_{21}}{6} \, , \label{eq:14b} \\
    k_{13} (4k_{11} - 9) &=& \frac{k_{11}}{4} - 2 k^2_{12} -
    \frac{k_{20}}{24} + \frac{k_{22}}{2} \, . \label{eq:1c}
\end{eqnarray}  
\end{subequations} 
We have then proven that Eq.~\eqref{eq:27} is a solution of
Eqs.~\eqref{eq:4} as long as the coefficients $k_{1j}$ solve the
algebraic system~\eqref{eq:14}.

One more step is necessary to show the relationship between the
coefficients $k_{1j}$ with a given scalar field potential $V(\phi)$,
as Eqs.~\eqref{eq:14} cannot be solved unless we get some extra
information about the coefficients $k_{2j}$. As we shall see now, the
values of $k_{2j}$ can be easily inferred in most cases by a proper
combination of the field derivatives of the potential $V(\phi)$.

To do so, we first assume that the de Sitter critical point
also corresponds to a given value of the scalar field that we denote by
$\phi_{\rm dS}$. The condition for the existence of this critical
point is:
\begin{subequations}
  \label{eq:26}
\begin{equation}
  \label{eq:26a}
  \lim_{\theta \to 0} \frac{y_1}{y} = - 2\sqrt{6} \lim_{\phi
    \to \phi_{\rm dS}} \frac{\partial_\phi V^{1/2}}{\kappa V^{1/2}} =
  - \frac{\sqrt{6}}{\kappa} \lim_{\phi \to \phi_{\rm
      dS}} \partial_\phi (\ln V) = 0 \, .
\end{equation}
This is nothing but a more formal rephrasing of the condition that
$k_{10}=0$. It is obvious that we must expect that the de Sitter point
corresponds to a critical point of the scalar field potential
[i.e. $(\partial_\phi V)(\phi_{\rm dS})=0$], but the important thing
here is that we will refer to the critical point in terms of the
logarithmic derivative of the potential, which will also allow
us to consider cases in which the critical point does not necessarily
correspond to a local maximum in the potential [see, for instance, the
case of $V(\phi) \propto \phi^n$ in Sec.~\ref{sec:case-ii:-k_11}
below]. In this sense, Eq.~\eqref{eq:26a} must be rather considered a
functional formula to find the de Sitter value $\phi_{\rm dS}$.

Once we have found $\phi_{\rm dS}$, we can now calculate the expansion
coefficients $k_{2j}$ for any given potential as
\begin{eqnarray}
  k_{20} &=&  \lim_{\theta \to 0} \frac{y_2}{y} = - 12
             \lim_{\phi \to \phi_{\rm dS}} \frac{\partial^2_\phi
             V^{1/2}}{\kappa^2 V^{1/2}} = \lim_{\phi \to \phi_{\rm
             dS}} \frac{3 (\partial_\phi U)^2 - 6 U \partial^2_\phi
             U}{\kappa^2 U^2} \, , \label{eq:26b} \\
 \frac{k_{21}}{k_{11}} &=& \lim_{\theta \to 0} \frac{y_2 - k_{20}
                           y}{y_1} = \lim_{\phi \to \phi_{\rm dS}}
                           \frac{12 \partial^2_\phi V^{1/2} + \kappa^2
                           k_{20} V^{1/2}}{2 \sqrt{6}
                           \kappa \partial_\phi V^{1/2}} = \lim_{\phi
                           \to \phi_{\rm dS}} \frac{- 3 (\partial_\phi
                           U)^2 +6 U \partial^2_\phi U + \kappa^2
                           k_{20} U^2}{\sqrt{6} \kappa U \partial_\phi
                           U} \, , \label{eq:26c} \\
k_{22} &=& k^2_{11} \lim_{\theta \to 0} \frac{y(y_2 - k_{20}
                           y - k_{21}y_1/k_{11})}{y^2_1} -
                            \frac{k_{20}}{8} - \frac{k_{12}
                            k_{21}}{k_{11}} \nonumber \\
         &=& - k^2_{11} \lim_{\phi \to \phi_{\rm dS}}
                           \frac{[12 \partial^2_\phi V^{1/2} +
             \kappa^2 k_{20}
              V^{1/2} - (k_{21}/k_{11}) 2 \sqrt{6}
             \kappa \partial_\phi V^{1/2}] V^{1/2}}{24 (\partial_\phi
             V^{1/2})^2} -
                            \frac{k_{20}}{8} - \frac{k_{12}
                            k_{21}}{k_{11}} \, , \nonumber \\
         &=& k^2_{11} \lim_{\phi \to \phi_{\rm dS}} \frac{3 (\partial_\phi
                           U)^2 - 6 U \partial^2_\phi U - \kappa^2
                           k_{20} U^2 + \sqrt{6} \kappa
             (k_{21}/k_{11}) U \partial_\phi U}{6 (\partial_\phi U)^2}
             - \frac{k_{20}}{8} - \frac{k_{12} k_{21}}{k_{11}} \,
             , \label{eq:26d}
\end{eqnarray}
\end{subequations}
\end{widetext}
where we also show the expressions in terms of the dimensionless
potential $U(\phi) = V(\phi)/M^4$, with $M^4$ the constant parameter
that in general denotes the energy scale in the
potential. Equations~\eqref{eq:26} allow the calculation up to the third
coefficient in the expansion of $y_2$, but it is not difficult to
imagine similar (and rather more cumbersome) expressions for higher-order coefficients. Notice that at the end Eqs.~\eqref{eq:14}
and~\eqref{eq:26} form together a very general recipe for the
determination of the coefficients $k_{1j}$ that can be applied to any
scalar field potential.

It can be seen that Eqs.~\eqref{eq:26} involve the use of rather
complicated combinations of the field derivatives of the scalar field
potential $V(\phi)$, and in this respect our method resembles the
calculation of the parameters in the slow-roll approximation. However,
a key difference is that in our approach we only require the
calculation of constant coefficients (i.e. $k_{1j}$ and $k_{2j}$),
rather than the calculation of scalar field functions. Once the
coefficients have been determined, all that is left is to use the
exact solution~\eqref{eq:27} to calculate the inflationary quantities.

As with any other series expansion, the inflationary dynamics arising
from complicated potentials may not be well described by the third-order system~\eqref{eq:26}, and higher-order approximations may be
necessary in such cases, which in turn would imply the calculation of
a larger number of expansion coefficients $k_{1j}$. Another limitation
comes from the fact that the series expansion~\eqref{eq:5} cannot deal
properly with local minima in the potential, at which the quantity
$\ln [V(\phi)]$ diverges. However, such limitation may not be too
troublesome as inflation is expected to end well before the scalar
field reaches any minimum in the potential.

\subsection{Comparison with the slow-roll approximation \label{sec:slow-roll-appr}}
Before proceeding further, we recall that the most-common ansatz to
solve the scalar field equations of motion is the so-called slow-roll
approximation\cite{Liddle:2000cg,*Mukhanov:2005sc}, which consists of
the two assumptions: $H^2 \simeq \kappa^2 V/3$ and $\ddot{\phi}
\simeq 0$. The two can actually be written, in terms of our dynamical
variables, as $y=1$ and $y_1 = 6 \tan(\theta/2)$, respectively. This
little exercise shows that the slow-roll approximation is not
appropriate for the dynamical system approach: slow-roll implies that
$\theta=0$, and the fixing of the expansion coefficients $k_{1j}$
(namely, $k_{11}=3$, $k_{12}=0$, and $k_{13}=1/4$) without any trace
in them of the physical parameters from the scalar field
model. Coincidentally, though, the slow-roll values are exactly those
of a quadratic scalar field potential, see
Sec.~\ref{sec:case-ii:-k_11} below. 

However, the slow-roll formula can be considered an approximation
to the true behavior of the solution nearby the point $\theta =0$, and
then at linear order we find that $y_1 \simeq 3 \theta$, which results
in $k_{11}=3$. The latter is a very special value, as it provokes
the disappearance of the first terms on the right-hand side in
Eqs.~\eqref{eq:6} and~\eqref{eq:9}, and then only the coefficients
$k_{12}$ and $k_{13}$ can carry on information from the scalar field models
into the inflationary quantities. Moreover, from Eq.~\eqref{eq:12} we
can see that $k_{11}=3$ corresponds to $k_{20}=0$ so that in the
linear analysis of Sec.~\ref{sec:crit-points-stab} the growing
eigenvalue disappears, $\omega_+=0$ [see Eqs.~\eqref{eq:12}
and~\eqref{eq:24}], but the growing relationship $y_1 = -\omega_-
\theta$ now reads $y_1 = 3 \theta$. Thus, the value $k_{11}=3$ is not
a generic prediction of inflationary single-field models, but rather
it should be interpreted as the value found from Eqs.~\eqref{eq:14}
whenever the true solution coincides with the slow-roll approximation.

\subsection{Comparison with Hubble flow equations}
One of the preferred perturbative methods to study inflationary
solutions and to make comparisons with
observations\cite{Ade:2015lrj,Vennin:2015vfa} is the so-called Hubble
flow equations (HFE)\cite{Hoffman:2000ue,*Kinney:2002qn}, even though
its limitations to represent the full landscape of inflationary solutions
have been well
documented\cite{Liddle:2003py,*Vennin:2014xta,Coone:2015fha}.

As shown in Ref.~\cite{Liddle:2003py}, the HFE can be regarded as a Taylor
expansion of the coefficients around $\phi=0$, which can be considered
as the initial point for the inflationary solution. The range of
potentials $V(\phi)$ of which the dynamics can be described by the HFE is
then quite restricted, but an extension of the method using Pad\'e
approximants\cite{Coone:2015fha} serves to include also the so-called
plateau potentials that seem to be preferred by
observations\cite{Ade:2015lrj}.

As we have remarked above, our perturbative method also relies on a
Taylor expansion, but this time of the logarithmic derivative of the
potential $V(\phi)$ in terms of the angular variable $\theta$ along an
inflationary trajectory. This is quite convenient because, as noticed
in Ref.\cite{Vennin:2015vfa}, the true potential driving the dynamics
of a scalar field in a homogeneous and isotropic universe is indeed
$\ln [V(\phi)]$. Moreover, the Taylor expansion~\eqref{eq:5} is always
(indirectly) calculated around a de Sitter critical point, represented
by $\theta=0$, for which the condition~\eqref{eq:26a} is
accomplished. The latter may correspond to a true critical point of the
potential, i.e. $[\partial_\phi V](\phi_{dS}) =0$ with $\phi_{dS} =
{\rm const.}$, or to the vanishing of the logarithmic derivative itself
$[\partial_\phi \ln(V)](\phi_{dS}) =0$ with $\phi_{dS} = \pm \infty$,
as usually happens in the case of monomial and plateau potentials
(see Sec.~\ref{sec:gener-class-solut} below). Thus, in contrast to the
HFE, there is no need in our method to consider different expansions
in order to deal with different types of potentials.

\subsection{Comparison with standard scalar field dynamics \label{sec:comp-with-stand}}
Strictly speaking, the best approach to find the predictions of a
given inflationary model is to solve directly the original equations
of motion~\eqref{eq:1}, and this is also the safest approach for
accurate enough
solutions\cite{Ade:2015lrj,Mortonson:2010er,Easther:2011yq}. To do so,
one has to be specific about the values of the potential parameters
and about the setup of initial conditions
($\phi_i,\dot{\phi}_i,H_i$). This procedure has to be made case by
case and only if the functional form of the scalar field potential is
known beforehand. Moreover, to use the observational constraints one
has to consider the three inflationary values $(A_S,n_S,r)$, as the
amplitude $A_S$ is required to put constraints on the energy scale
$M^4$ of the scalar field potential.

In contrast, in our perturbative approach the general properties of
the inflationary solutions can be explored in full generality, as
Eq.~\eqref{eq:27} is the only solution for all possible cases, whether
we know them specifically or not. In principle, it would suffice to
use the observational values of only $(n_S,r)$ to constrain the
coefficients $k_{11}$, $k_{12}$ and $k_{13}$. A complete comparison with
observations is beyond the purposes of this paper, but a glimpse of it
will be given in Sec.~\ref{sec:comp-with-observ} below. Notice that
the $A_S$ is not required for the task, simply because the energy
scale of the scalar field potential will never be present in the
determination of the coefficients $k_{1j}$. This is an indication that
$A_S$ and $(n_S,r)$ put constraints on different targets, and then it
would be wiser to use them separately, that is, to use $(n_S,r)$ to
constrain the internal physical parameters of the scalar field
potential, and later use $A_S$ to constrain the potential energy
scale. Unfortunately, such a separation cannot be made in the
full numerical approach.

Another advantage is that Eqs.~\eqref{eq:4} are certainly more
manageable if one wishes to find the background dynamics
numerically. First of all, we only require determining the functional
form of $y_2$; this can either be done exactly for some selected cases
(like the ones considered in Sec.~\ref{sec:gener-class-solut} below)
or approximately by taking the truncated series expansion $y_2 \simeq k_{20}
+ k_{21}\theta + k_{22} \theta^2 + \ldots$, once the coefficients
$k_{2j}$ are calculated from Eqs.~\eqref{eq:14} and~\eqref{eq:26}. In
the second place, the setup of initial conditions is already given by the
assumption that we depart from a de Sitter critical point: $y_{1i} =
k_{11}\theta_i$, where $\theta_i$ can be estimated as a first guess
from the general third-order solution~\eqref{eq:27} as $\theta_i
\simeq \theta_N$ for any desired number $N$ of $e$-folds before the
end of inflation. Such numerical solution of the background would then
be used to solve separately the equations of motion of scalar and
tensor perturbations (see, for instance,
Ref.~\cite{Urena-Lopez:2015gur} for a dynamical system approach to the
evolution of scalar field density perturbations).

\section{Universal classes of solutions and accuracy tests\label{sec:gener-class-solut}}

To test the accuracy of the analytical solution~\eqref{eq:27}, we will
make some comparisons with the numerical solutions obtained from the
full equations of motion~\eqref{eq:4} in selected cases that do not
require an approximate solution, which means that the variable $y_2$
can be calculated exactly. 

We will take the nonperturbative expression of $y_2$ that
corresponds to those selected models, and substitute it on the rhs of
Eq.~\eqref{eq:4b}. The value of $\theta_N$ will be obtained
numerically as a function of $N$, and then used to calculate the
inflationary quantities through Eqs.~\eqref{eq:8}. In contrast, for
the perturbative analytical solution, we will calculate the
coefficients $k_{11}$, $k_{12}$ and $k_{13}$ for the given potentials and
determine the value of $\theta_N$ from Eq.~\eqref{eq:27}. The
obtained values will be substituted into the (perturbative)
Eqs.~\eqref{eq:9} of the inflationary quantities.

The relative errors of $n_S$ and $r$, that is, of their perturbative
values with respect to their exact numerical ones, will be plotted
as a function of the number $N$ of $e$-folds before the end of
inflation. To ease the comparison, and for reasons that will be
explained better in Sec.~\ref{sec:comp-with-observ} below, we will
focus our attention in two general classes of solutions in terms of
the coefficient $k_{12}$.

\subsection{Class I: $k_{12}= 0$ \label{sec:case-ii:-k_11}}
This is a case that deviates from the slow-roll prediction, as it
corresponds to $k_{20} \neq 0$, even though still $k_{21}=0$. The
solution can be obtained from the general expression~\eqref{eq:27b} in
the limit $k_{12} \to 0$, and then we can write:
\begin{equation}
  \theta_N = \theta_{end} |\theta_\pm| \left[ \left( |\theta_\pm|^2 +
      \theta^2_{end} \right) e^{2(k_{11}-3)N} - \theta^2_{end}
  \right]^{-1/2} \, , \label{eq:18}
\end{equation}
where $|\theta_\pm|^2 = 2(k_{11}-3)/(2k_{13}+1)$, see
Eq.~\eqref{eq:27c}.

An example here is \emph{natural inflation} (NI)\cite{Freese:2014nla,*Freese:1990rb,*Adams:1992bn}. The scalar field
potential is $V(\phi)= M^4[1+\cos(\phi/f)]$, for which we find:
\begin{subequations}
\label{eq:19}
  \begin{eqnarray}
    V^{1/2} &=& \sqrt{2} M^2 \cos(\phi/2f) \, , \\
    \partial_\phi V^{1/2} &=& - \frac{1}{\sqrt{2}} \frac{M^2}{f}
    \sin(\phi/2f) \, , \\
    \partial^2_\phi V^{1/2} &=& - \frac{1}{2\sqrt{2}}
    \frac{M^2}{\kappa f^2}
    \cos(\phi/2f) \, .
  \end{eqnarray}
\end{subequations}
The combination of the field derivatives~\eqref{eq:19} results in
the relation $y_2 = (3/\kappa^2 f^2) y$. In terms of the expansion
coefficients~\eqref{eq:5}, we find that $k_{20}=3/(\kappa^2 f^2)$,
$k_{21}=0$, and $k_{22}=-3/(8\kappa^2 f^2)$. Notice that these same
results can also be obtained, although with more tiresome
calculations, using Eqs.~\eqref{eq:26} with $\phi_{\rm dS} =0$. From
Eqs.~\eqref{eq:14} we obtain $k_{12}=0$, and also:
\begin{subequations}
\label{eq:20}
  \begin{eqnarray}
    k_{11} &=& \frac{3}{2} \left( \sqrt{1+ \frac{2}{3}
        \frac{1}{\kappa^2 f^2}}
    + 1 \right) \, , \\
    k_{13} &=& \frac{1}{4k_{11}-9} \left( \frac{k_{11}}{4} -
      \frac{5}{16} \frac{1}{\kappa^2 f^2} \right) \, .
  \end{eqnarray}
\end{subequations}

The substitution of Eqs.~\eqref{eq:20} into Eq.~\eqref{eq:18} provides
an almost exact solution of NI for any value of the decay constant
$f$, which is in agreement with others reported in the literature that
go beyond the slow-roll
approximation\cite{Martin:2014vha}.\footnote{Another related potential
  is the so-called \emph{hybrid natural inflation} (HNI), in which the
  potential is generalized to $V(\phi)= M^4[1+a\cos(\phi/f)]$, where
  $0 < a < 1$ is a constant\cite{Ross:2016hyb,*Vazquez:2014uca}. If we
  apply our perturbative formalism, it can be shown that $k_{20} = -8
  k_{22}= (3/\kappa^2 f^2)[2a/(1+a)]$ and $k_{21}=0$, which indicates
  that HNI still belongs to Class I despite the presence of an extra
  parameter. Notice that for the calculation of $(n_S,r)$, HNI provides
  the same results as NI but with a rescaled decay constant $\hat{f}^2 =
  f^2 (1+a)/2a$.} In the limit $\kappa f \gg 1$ we recover from
Eq.~\eqref{eq:18}, as expected, the case of the quadratic potential
(see below).

\begin{figure}[htp!]
\centering
\includegraphics[width=0.49\textwidth]{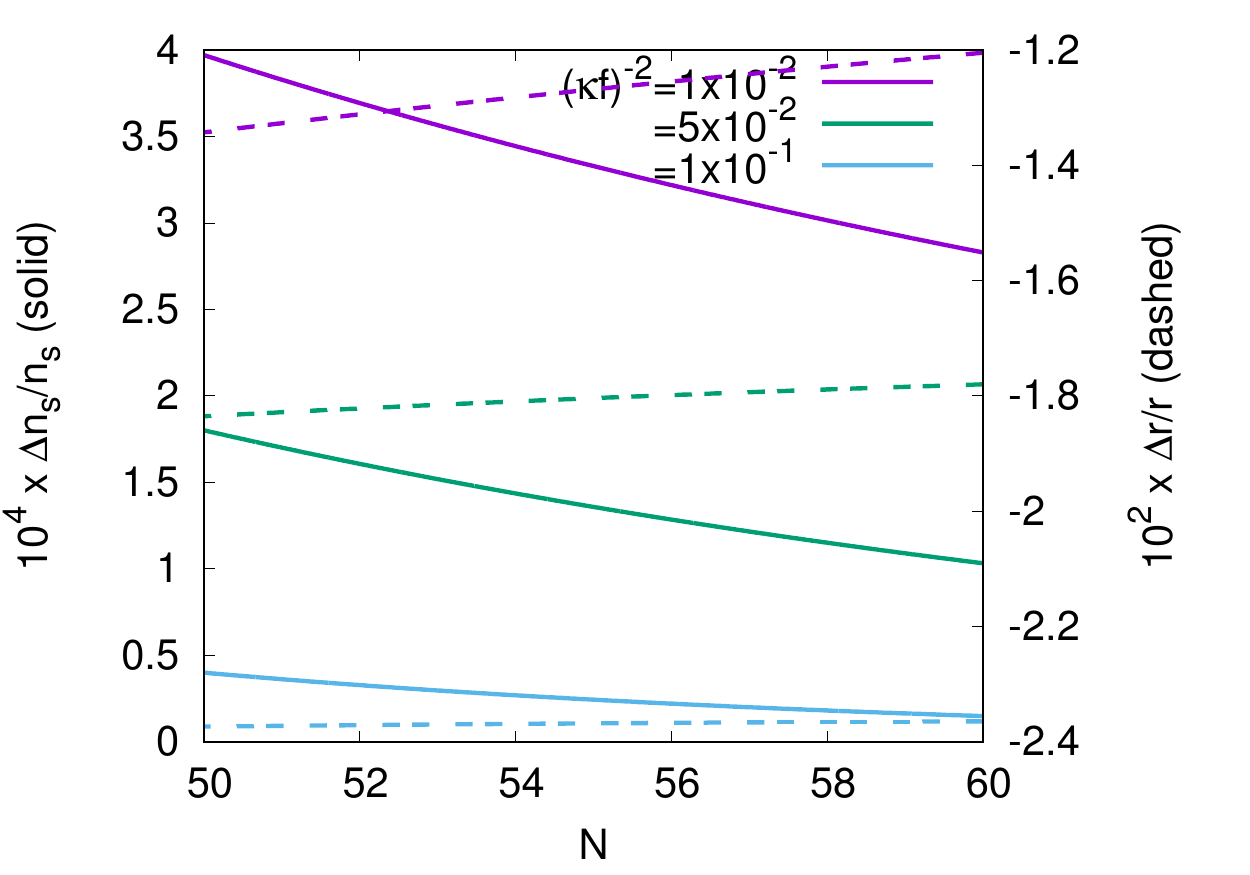}
\caption{\label{fig:0} Relative errors in the values of inflationary
  quantities in case of NI, see Eqs.~\eqref{eq:19}, as calculated from
  the analytical solution~\eqref{eq:18} and the numerical one obtained
  from Eqs.~\eqref{eq:4}. The curves are labeled in terms of the
  dimensionless quantity $(\kappa f)^{-2}$, where $f$ is the so-called
  decay constant. The plotted values correspond to the formula $
  (g^{pert} -g^{num})/g^{num}$, where $g$ can either be the spectral
  index $n_S$ (left vertical axis, solid curves) or the
  tensor-to-scalar ratio $r$ (right vertical axis, dashed curves). The
  relative errors are less than $0.04\%$ ($2.4\%$) in the case of $n_S$
  ($r$) for the values of interest at $N=50-60$ $e$-folds before the
  end of inflation.}
\end{figure}

The comparison between the numerical and the perturbative inflationary
solutions in the case of NI is shown in Fig.~\ref{fig:0}. The relative
errors of the inflationary quantities is below $0.04\%$ ($2.4\%$) for the
spectral index $n_S$ (the tensor-to-scalar ratio $r$).

To finish this section, we now show that the simplest inflationary
case allowed by our perturbative approach belongs to Class I, and can
be obtained from Eq.~\eqref{eq:18} in the limit $k_{11} \to 3^+$:
\begin{equation}
  \theta_N = \theta_{end} \left[ (2k_{13}+1) \theta^2_{end} N
  + 1 \right]^{-1/2} \, . \label{eq:15}
\end{equation}

One typical example in this case is \emph{large field inflation}
  (LFI). The scalar field potential is $V(\phi)= M^4(\phi/\phi_0)^{2n}$\cite{Linde:1983gd}, for which we find the
following field derivatives:
\begin{subequations}
\label{eq:16}
  \begin{eqnarray}
    V^{1/2} &=& M^2 (\phi/\phi_0)^n \, , \\
    \partial_\phi V^{1/2} &=& n M^2 \phi^{n-1}/\phi^n_0 \, , \\
    \partial^2_\phi V^{1/2} &=& n(n-1) M^2 \phi^{n-2}/\phi^n_0 \, .
  \end{eqnarray}
\end{subequations}
In terms of the definitions~\eqref{eq:2b}, Eqs.~\eqref{eq:16} can be
combined as $y_2 = [(1-n)/2n] y^2_1/y$. Writing the latter expression
in terms of the expansion coefficients~\eqref{eq:5}, we find that
$k_{20}=0 = k_{21}$ and $k_{22}=[(1-n)/2n] k^2_{11}$. As before, the
same results can be obtained from Eqs.~\eqref{eq:26} with $\phi_{\rm
  dS} = -\infty$, which is the critical value of the scalar field
indicated by the condition~\eqref{eq:26a}.

If we now use Eqs.~\eqref{eq:14}, we obtain that $k_{11}=3$,
$k_{12}=0$, and $k_{13} = 3/4n - 1/2$. Notice that $k_{13}$ is
negative for $n>1$, but the important combination in Eq.~\eqref{eq:6}
is $(k_{13}+1/2)$, which is always positive. If we substitute these
values in Eqs.~\eqref{eq:9} and~\eqref{eq:15}, we find the exact results\cite{Liddle:2000cg,Martin:2014vha}:
\begin{equation}
  \label{eq:25}
  1 - n_S = \frac{3(n+1)}{3N+2n\theta^{-2}_{end}} \, , \quad r =
  \frac{24n}{3N+2 n\theta^{-2}_{end}} \, .
\end{equation}
In the limit $n \ll 1$ we find that the predictions are $1-n_S = 1/N$
and $r=0$; see also Eq.~\eqref{eq:29} in Sec.~\ref{sec:class-i-}
below.

\begin{figure}[htp!]
\centering
\includegraphics[width=0.49\textwidth]{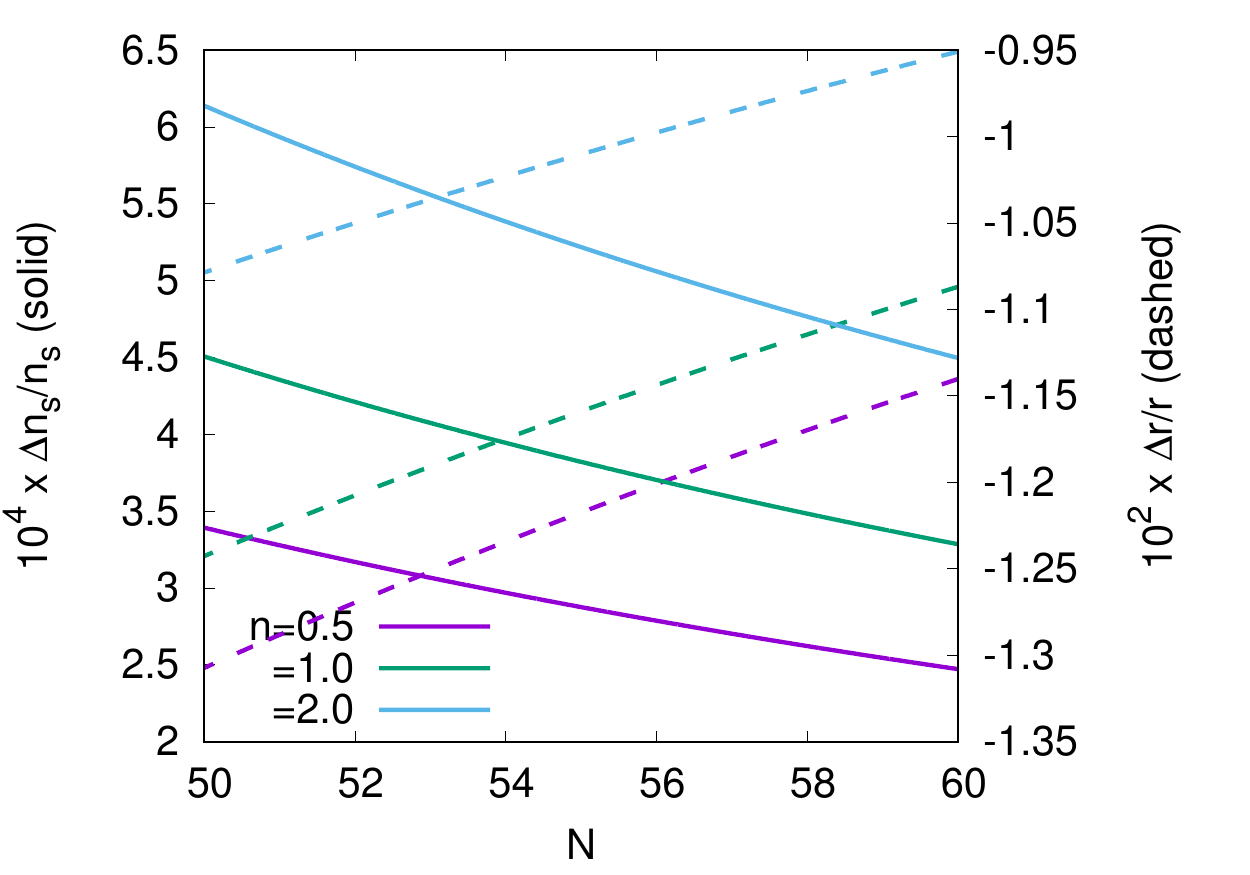}
\caption{\label{fig:2} The same as in Fig.~\ref{fig:0}, but now for
  the case of LFI, see Eqs.~\eqref{eq:25}. The relative errors between
  the approximate solution~\eqref{eq:15} and the exact numerical one
  is less than $0.06\%$ ($1.3\%$) for $n_S$ ($r$) for the values of
  interest at $N=50-60$ $e$-folds before the end of inflation.}
\end{figure}

The comparison between the exact numerical solution and the
perturbative inflationary one in Eq.~\eqref{eq:25}, in the case of
LFI, is shown in Fig.~\ref{fig:2}. The relative errors of the
inflationary quantities are below $0.06\%$ ($1.3\%$) for the spectral
index $n_S$ (the tensor-to-scalar ratio $r$).

\subsection{Class II: $k_{12} \neq 0$ \label{sec:case-iv:-k_11}}
In principle, this class is defined simply as all inflationary cases
of which the solution is given by Eqs.~\eqref{eq:27} but that do not belong
to Class I, which means that $k_{11} \geq 3$, both $k_{12}$ and $k_{13}$ are
non-zero, and all three coefficients contribute to the inflationary
solution. Unfortunately, there does not seem to be a clear
representative example of this general case.

But there is indeed, as in Class I, a particular subclass of solutions
within Class II that deserves a separate study. The second- and
third-order solutions are
\begin{subequations}
\label{eq:17}
\begin{eqnarray}
  \theta^{(2nd)}_N &=& \theta_{end} \left( \theta_{end} k_{12} N +1
  \right)^{-1} \, , \label{eq:17a} \\
  k_{12} N^{(3rd)} &=& \frac{1}{\theta_N} - \frac{1}{\theta_{end}} -
  \frac{1}{\theta_-} \ln \left( \frac{1 - \theta_-/\theta_{end}}{1 -
      \theta_-/ \theta_N} \right) \, , \label{eq:17b}
\end{eqnarray}
\end{subequations}
where $\theta_- = -2k_{12}/(2k_{13} +1)$ (and also $\theta_+ =0$). It
can be easily shown that Eqs.~\eqref{eq:17} can be obtained from the
general solution~\eqref{eq:27} in the limit $k_{11} \to 3^+$.

Examples of this subclass are the \emph{Starobinsky's
  model}\cite{Starobinsky:1980te} and, more generally, the so-called
$\alpha$ \emph{attractors}\cite{Carrasco:2015pla,*Kallosh:2013yoa,*Galante:2014ifa}
(see also the Higgs Inflation model in\cite{Martin:2014vha}). For
general purposes of illustration, we write the scalar field potential
as $V(\phi)= M^4(1- e^{-\lambda \kappa \phi})^2$, for which we find
\begin{subequations}
\label{eq:21}
  \begin{eqnarray}
    V^{1/2} &=& M^2 \left( 1 - e^{\lambda \kappa \phi} \right) \, , \\
    \partial_\phi V^{1/2} &=& -\lambda \kappa M^2 e^{\lambda \kappa
      \phi} \, , \\
    \partial^2_\phi V^{1/2} &=& -\lambda^2 \kappa^2 M^2 e^{\lambda \kappa
      \phi} \, ,
  \end{eqnarray}
\end{subequations}
where we have changed the conventional signs so that the perturbative
solution~\eqref{eq:6} picks up the correct branch in the potential
that corresponds to the de Sitter solution. Otherwise, we would have
encountered the power-law solution related to the second critical
point in the dynamical system, see
Sec.~\ref{sec:crit-points-stab}. 

Equations~\eqref{eq:21} suggest that $y_2 = \sqrt{6} \lambda y_1$, which in turn
means that $k_{20}=0$, $k_{21}= \sqrt{6} \lambda k_{11}$, and
$k_{22}= \sqrt{6} \lambda k_{12}$. From Eqs.~\eqref{eq:14} we find
that $k_{11}=3$, $k_{12} = (\sqrt{6}/2) \lambda$, and $k_{13}= 1/4 -
\lambda^2/2$. Needless to say, the same values are also obtained from
Eqs.~\eqref{eq:26} for the de Sitter critical point $\phi_{\rm dS} =
-\infty$.

Even though Eq.~\eqref{eq:17b} is an analytical third-order
solution, the one considered in most papers is the second-order one in
Eq.~\eqref{eq:17a}\cite{Martin:2014vha}. For the latter, in the large
$N$ limit, we find that $\theta_N \simeq (1/k_{12}) N^{-1} =
(2/\sqrt{6} \lambda) N^{-1}$. It is also true that in the limit
$\lambda \to 0$ the expansion coefficients are $k_{11}=3$, $k_{12}
=0$, and $k_{13}= 1/4$, which corresponds to the quadratic potential
(see Sec.~\ref{sec:case-ii:-k_11} above). These two limiting behaviors
of this type of model have been widely studied in the
literature\cite{Galante:2014ifa}.

The parameter $\theta_-$ diverges for $\lambda^2=3/2$, and then we
find that, in the limit $\lambda \to \sqrt{3/2}$, Eq.~\eqref{eq:17b}
transforms into Eq.~\eqref{eq:17a}. We recall here that the original
Starobinsky model corresponds to $\lambda =
\sqrt{2/3}$\cite{Starobinsky:1980te,Ade:2015lrj}, and then its
solution is well represented by Eq.~\eqref{eq:17a} with
$k_{12}=1$. For values $\lambda^2 > 3/2$ the parameter $\theta_-$
becomes negative, and then Eq.~\eqref{eq:17b} does not give consistent
solutions for all cases, which may indicate the need to consider
higher-order corrections in the inflationary solution~\eqref{eq:6}.

\begin{figure}[htp!]
\centering
\includegraphics[width=0.49\textwidth]{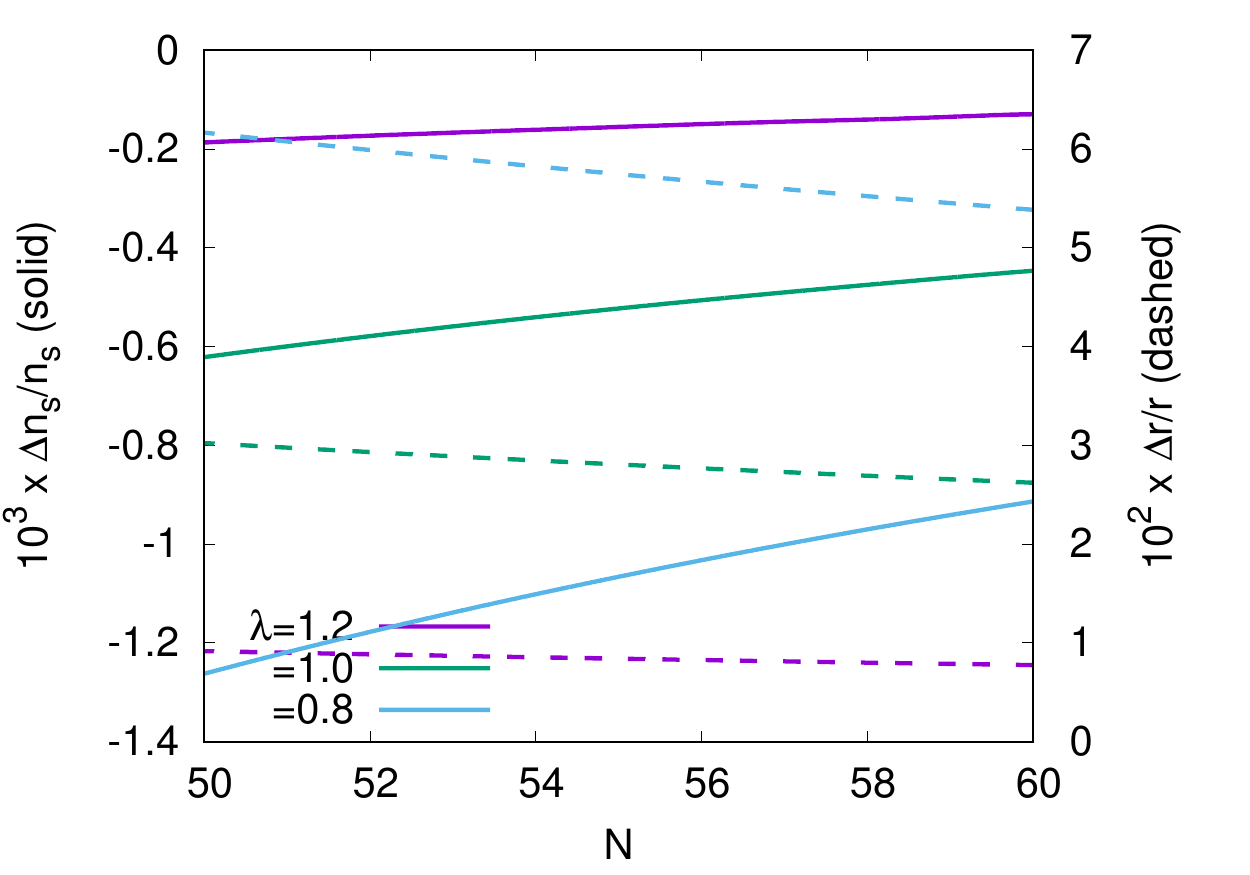}
\caption{\label{fig:4} The same as in Figs.~\ref{fig:0}
  and~\ref{fig:2}, but now for the case of the Starobinsky's model,
  see Eqs.~\eqref{eq:21}. The perturbative solutions were calculated
  from the second-order equation~\eqref{eq:17a}. The curves are labeled in
  terms of the parameter $\lambda$ that appears in the potential. The
  relative errors are less than $0.14\%$ ($6\%$) in the case of $n_S$
  ($r$) for the values of interest at $N=50-60$ $e$-folds before the
  end of inflation.}
\end{figure}

As we did before for the case of NI, the comparison between the
numerical and the perturbative inflationary solutions in the case of
the Starobinsky model is shown in Fig.~\ref{fig:4}. The relative
errors of the inflationary quantities is below $0.14\%$ ($6\%$) for
the spectral index $n_S$ (the tensor-to-scalar ratio $r$). Notice
that for simplicity the comparison was made in terms of the second-order perturbative solution in Eq.~\eqref{eq:17a}, but a better
result can be obtained if the third-order one in Eq.~\eqref{eq:17b} is
used instead.

\section{General results on the $n_S-r$ plane \label{sec:comp-with-observ}}
In the above sections we discussed the mathematical properties of the
general solutions arising from single models of inflation. Those
solutions are characterized by three numbers: $k_{11}$, $k_{12}$ and
$k_{13}$, and these same numbers are the ones that determine the
values of the observables, so that $n_S=n_S (k_{11}, k_{12},
k_{13})$, and $r=r (k_{11}, k_{12}, k_{13})$.

The two classes presented in Sec.~\ref{sec:gener-class-solut} were
each exemplified each one by typical cases that are well known in the
literature, but as we shall see now, the capabilities of the
inflationary models are more extended than those suggested by the
given instances of particular potentials. 

The calculations below were made under the assumption that
the numbers $k_{11}$, $k_{12}$ and $k_{13}$ are all independent. As we saw
before, Eqs.~\eqref{eq:14} and~\eqref{eq:26} indicate that they are
not completely independent once we choose a particular potential
$V(\phi)$, but here we will take a less restrictive point of view,
because we only want to make a estimation of the values of $k_{11}$,
 $k_{12}$ and $k_{13}$ that are in principle compatible with observations.

\subsection{Class I \label{sec:class-i-}}
\begin{figure*}[htp!]
\centering
\includegraphics[width=0.49\textwidth]{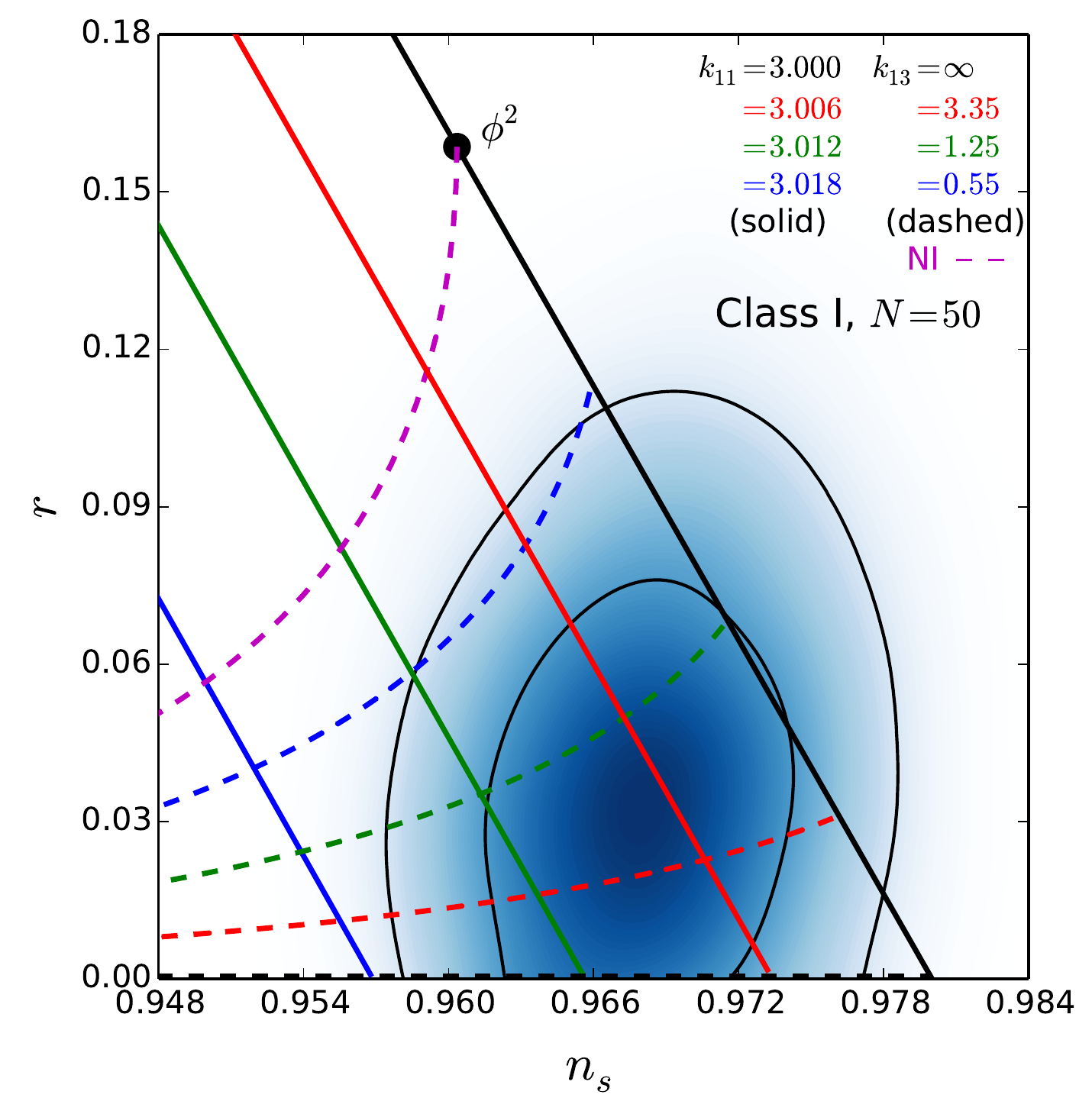}
\includegraphics[width=0.49\textwidth]{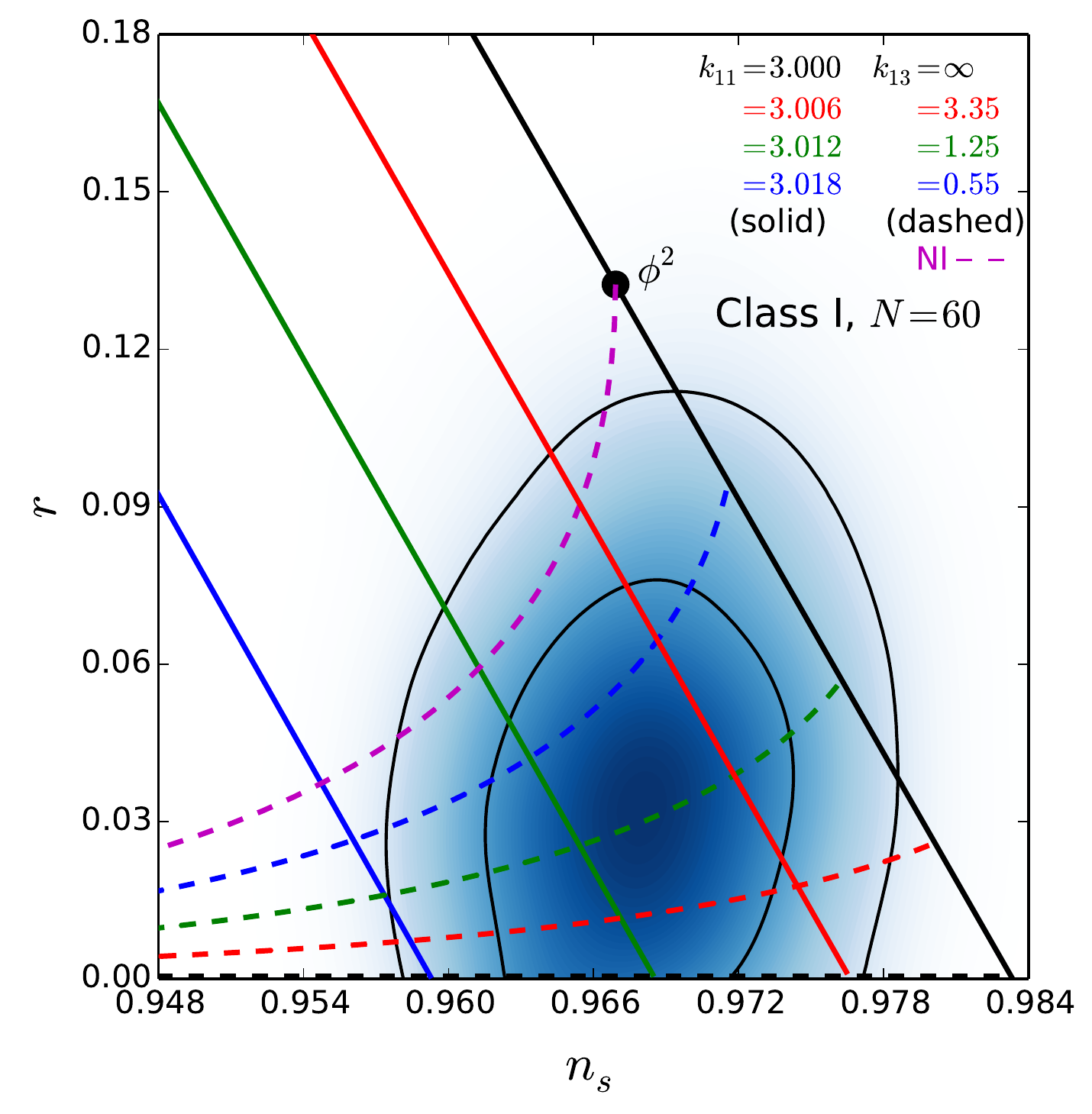}
\caption{\label{fig:1} General results on the plane $n_S$ vs $r$ for
  the Class I, from Sec.~\ref{sec:case-ii:-k_11}, of single-field
  models of inflation. The plots show the results for $N=50$ (left)
  and $N=60$ (right). Notice that Class I is not completely ruled out
  by observations, mostly because it can provide low enough values of
  the tensor-to-scalar ratio $r$. The predictions from the typical LFI
  for the potential $\phi^2$ [$n=1$ in Eq.~\eqref{eq:16}], and NI
  [Eq.~\eqref{eq:19}], are indicated in the figures for
  comparison. The contours represent the $65$\% and $95$\% confidence
  levels of the {\it Planck} TT+lowP+BAO data\cite{Ade:2015lrj}.}
\end{figure*}

The predictions for the plane $n_S-r$ are shown in Fig.~\ref{fig:1}
for $N=50,60$ $e$-folds before the end of inflation. We can see that
if we keep $k_{11} = {\rm const}$, the resultant plots are straight
lines which are parallel to the one corresponding to $k_{11}=3$. As
expected from Eq.~\eqref{eq:9}, this confirms that $k_{11}$ has a
major effect on the value of the spectral index $n_S$, and then a good
fit to the observational data requires that $3.018 > k_{11} >
3$. This also shows that the value of $k_{11}$ should be away from
the slow-roll value $k_{11}=3$, but not by much. On the other hand,
the curves that are obtained from $k_{13} = {\rm const.}$ are not
straight lines, and they show that $k_{13}$ has a major effect in the
value of the tensor-to-scalar ratio $r$, and that in principle $k_{13}
> 0.5$ is required in order to obtain a low enough value of $r$.

Actually, it can be shown that in the limit $k_{13} \to \infty$ of
Eq.~\eqref{eq:18}, which should result in a null value of $r$, the
prediction for the spectral index is:
\begin{equation}
  \label{eq:29}
  1-n_S \simeq 2(k_{11}-3) + 1/N \, .
\end{equation}
In the calculation of Eq.~\eqref{eq:29} we have also considered that
$(k_{11}-3)N \ll 1$, which seems to be approximately correct in the
cases of interest; see Fig.~\ref{fig:1}. Equation~\eqref{eq:29} also
suggests that for models in Class I there is an upper bound in the
value of the spectral index given by $n_S \leq 1-1/N$; this results in
$n_S \leq 0.98$ if $N=50$ ($n_S \leq 0.983$ if $N=60$).

Just for comparison, we also show in Fig.~\ref{fig:1} the inflationary
results obtained from the LFI ($\phi^2$) and the NI potentials. As it
is now widely accepted, those potentials seem to be ruled out by
observations\cite{Martin:2014vha}. Notice in particular that the curve
from the NI potential almost corresponds to an isoline of $k_{13} =
{\rm const.}$, but the value of the latter is not large enough to be
in agreement with the observational constraints.

\subsection{Class II \label{sec:class-ii-}}
\begin{figure*}[htp!]
\centering
\includegraphics[width=0.49\textwidth]{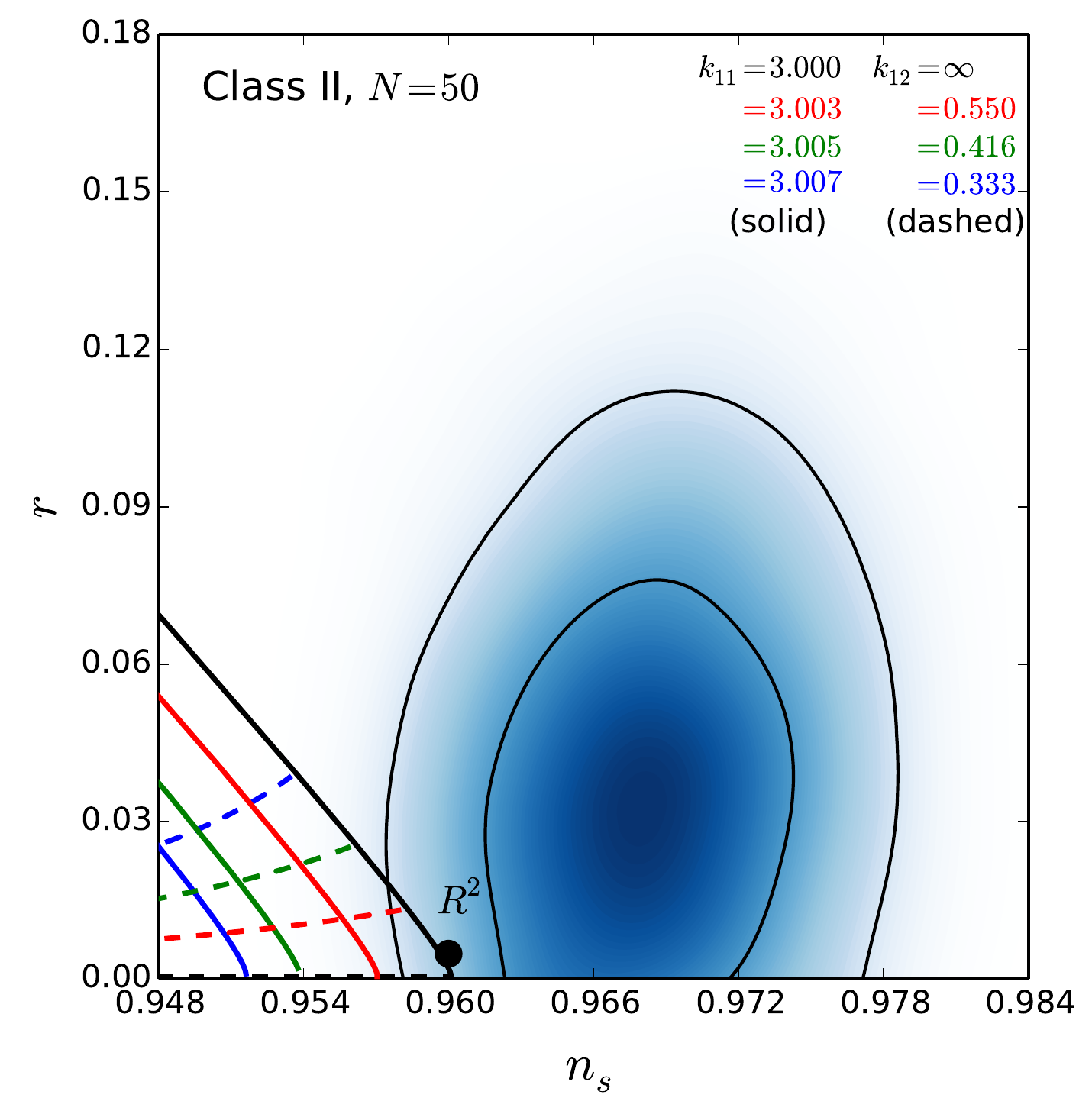}
\includegraphics[width=0.49\textwidth]{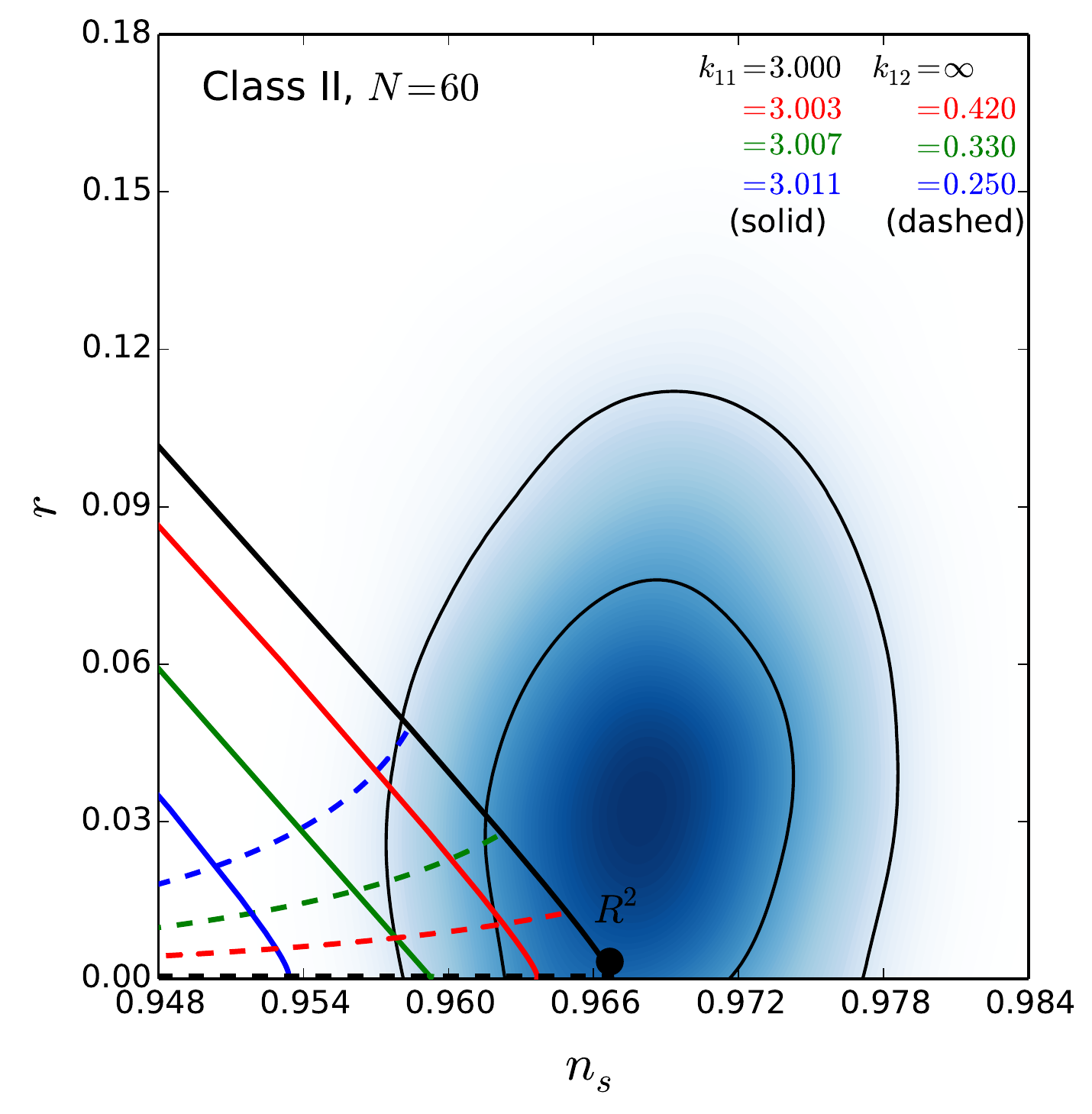}
\caption{\label{fig:5} General results on the plane $n_S$ vs $r$ for
  Class II, from Sec.~\ref{sec:case-iv:-k_11}, of single field models
  of inflation. The plots show the results for $N=50$ (Left) and
  $N=60$ (Right). Notice that Class II covers a smaller area than
  Class I (see Fig.~\ref{fig:1}), and then it has less freedom to fit
  the observational data. The predictions from the typical
  Starobinsky's model ($R^2$) ($\lambda=\sqrt{2/3}$ in
  Eq.~\eqref{eq:21}), is indicated in the figures for comparison. The
  contours represent the $65$\% and $95$\% confidence levels of the
  {\it Planck} TT+lowP+BAO data\cite{Ade:2015lrj}.}
\end{figure*}

For the comparison of this class of inflationary solutions, we will
only consider the second-order solutions in Eqs.~\eqref{eq:27a}
and~\eqref{eq:17a}. Although the third-order solutions would be more
precise, the second-order ones are accurate enough for the purposes of
this section, and they will allow us to show the main results in a
similar way as for Class I above. The general predictions from the
models in Class II are presented in Fig.~\ref{fig:5}.

If we keep $k_{11} = {\rm const.}$, the resultant plots are straight lines
which are parallel to the one corresponding to $k_{11}=3$. This again
shows that $k_{11}$ has a major effect on the value of the spectral
index $n_S$. However, a major difference with respect to Class I
appears here: the region covered by Class II in the parameter space
moves away considerably from the constrained region if $k_{11} >
3$. Hence, the best option here seems to be the slow-roll value
$k_{11}=3$. On the other hand, the curves that are obtained from
$k_{12} = {\rm const.}$ show that $k_{12}$ has a major effect on the
value of the tensor-to-scalar ratio $r$ (similar to the role of
$k_{13}$ in Class I) and that in principle $k_{12} > 0.2$ is required
in order to obtain a low enough value of $r$.

A difference with respect to the role of $k_{13}$ in Class I, is that the
value of $r$ vanishes more quickly as $k_{12}$ grows, which is explained
from the fact that for models in Class II we have that $\theta_N \sim
N^{-1}$, whereas for those in Class I the result is just $\theta_N \sim
N^{-1/2}$. This also has a visible effect on the spectral index. If we
take the limit $k_{12} \to \infty$ of Eq.~\eqref{eq:27a}, we find that
\begin{equation}
  \label{eq:30}
    1-n_S \simeq 2(k_{11}-3) + 2/N \, .
\end{equation}
In the calculation of Eq.~\eqref{eq:30} we have again considered that
$(k_{11}-3)N \ll 1$, which seems to be approximately correct in the
cases of interest, see Fig.~\ref{fig:5}. Equation~\eqref{eq:30} also
suggests that for models in Class II there is an upper bound in the
value of the spectral index given by $n_S \leq 1-2/N$; this results in
$n_S \leq 0.96$ if $N=50$ ($n_S \leq 0.966$ if $N=60$). This in turn
indicates that models in Class II can only explore a smaller range of
values of $n_S$ than those in Class I, and this limitation may put
them in jeopardy when compared with observations.

\section{Conclusions \label{sec:conclusions}}
We have presented a new transformation of the equations of motion in
single-field inflationary models that render them in a more suitable
form for a dynamical system analysis than in other standard
approaches. The evolution of the scalar field is given mostly by an
internal angle variable $\theta$, which together with other potential
variables, allows a perturbative solution around a critical, de Sitter,
point at any desirable order.

It must be remarked that the perturbative method does not require the
imposition of the slow-roll approximation, and can actually be applied
to scalar field models even if the slow-roll conditions are not
strictly attained. Also, in contrast to slow roll, the method does not
explicitly need to resolve the evolution of the scalar field, and the
full solution is just represented by the angular variable
$\theta$. Given that $w_\phi =- \cos \theta$ (see Sec.~\ref{sec:scal-field-dynam}),
this illustrates the fact that, for inflationary calculations, the
evolution of the equation of state $w_\phi$ is all that is required
for a full inflationary solution.

In addition, the new equations of motion showed that there are
two critical points of physical interest, which in turn suggested
that the inflationary solutions appear in two flavors. The first
critical point corresponds to power-law inflation, which only
appears for a very restricted type of exponential-like
potentials. The second critical point corresponds to de Sitter
inflation, which is a generic case for a broad range of scalar field
potentials, and this is why its solutions were studied more in
detail than those of the power-law one.

The main source of error in the inflationary solutions comes precisely
from the determination of $\theta_N$ itself, the value of the angular
variable at a given number of $e$-folds before the end of
inflation. The value of $\theta_N$ is mainly affected by the
truncation in the expansion of $y_1$, but we showed that at least in
some selected examples the accuracy of the method is good
enough. Furthermore, the effect of $\theta_N$ is less important in the
determination of $n_S$, and then the predictions for the spectral
index have a much larger accuracy than those for the tensor-to-scalar
ratio $r$.

The new method suggests that de Sitter inflation in single-field
models can be arranged in two general classes, which we called Class
I and Class II, irrespective of the considered scalar field
potential. We were able to identify representative examples for each
class, although an example that fully shows the features of Class II
seems not to have been reported before. Interestingly enough, the
key parameter for such a classification is the coefficient
$k_{12}$. If we take the large $N$ limit of the two classes, we end
up with only two typical behaviors in the solutions, namely,
$\theta_N \sim N^{-1/2}$ ($k_{12}=0$: Class I) or $\theta_N \sim
N^{-1}$ ($k_{12} \neq 0$: Class II), which suggests that either $r
\sim N^{-1}$ or $r \sim N^{-2}$, respectively, even though $1-n_S
\sim N^{-1}$ at the leading order in $N$ for the two classes. This
is in agreement with recent works that suggest the existence of these
\emph{universal} classes in
inflation\cite{Roest:2013fha,*Galante:2014ifa} (see
also Ref.~\cite{Mukhanov:2013tua}). As for the comparison with observations,
it seems that models in Classes I are the most suitable to fit the
current constraints because of the suppressed value of the
tensor-to-scalar ratio $r$, and the more extended freedom to fit the
preferred value of the spectral index $n_S \simeq
0.965$\cite{Ade:2015lrj} [see Eq.~\eqref{eq:29} above].

The dynamical treatment presented here also showed that the
determination of the inflationary quantities $(n_S,r)$ is insensitive
to the energy scale of the scalar field potential, and then the latter
may be left out in any fitting analysis. This is not possible, though,
in the standard approach to scalar field dynamics, in which none of
the potential parameters can be avoided in the calculation of the
inflationary trajectories.

A key ingredient in the method was the transformation of the dynamical
system into a hierarchy of algebraic relations for the numerical
coefficients in a series expansion. We showed that there is a
shortcut to solving such an algebraic system for any scalar field
potential with a given relation among its first potential variables in
the form $f(y,y_1,y_2) =0$. This is a relation between the scalar
field potential $V(\phi)$ and its first two derivatives through the
definitions~\eqref{eq:2b}. The examples we considered above cover some of
the most popular instances in recent inflationary studies;
see Refs.~\cite{Ade:2015lrj,Martin:2014vha,Cook:2015vqa,*Cai:2015soa,*Munoz:2014eqa}
and references therein. Nonetheless, we showed that there is a general
procedure to calculate the expansion coefficients $k_{1j}$ of the
potential variable $y_1$, as long as one is able to identify the de
Sitter critical point in the given scalar field potential. At the end,
the method can be applied to a large variety of models, the
inflationary solutions of which would belong to those of either Class I or
Class II.

The framework presented here can in principle be extended
to more general situations, in which the scalar field would be endowed
to more complicated potentials. The hierarchy of algebraic
equations~\eqref{eq:14} and~\eqref{eq:26} will correspondingly become
more involved too, but this will not have any effect in our general
classification of inflationary cases. The reason is that such
classification only depends upon the given value of a single
coefficient ($k_{12}$), whatever further complications may arise in
the solution of the algebraic hierarchy.

\begin{acknowledgments}
I wish to thank Andrew Liddle and the Royal Observatory, Edinburgh,
for their kind hospitality during a fruitful sabbatical stay during
which part of this work was done. I am grateful to Eric Linder,
Vincent Vennin, and Alberto Diez-Tejedor for useful comments and
suggestions, and to Alma Gonz\'alez-Morales for help with redrawing
some figures. This work was partially supported by Programa para el
Desarrollo Profesinoal Docente; Direcci\'on de Apoyo a la
Investigaci\'on y al Posgrado, Universidad de Guanajuato, research Grant
No. 732/2016; Programa Integral de Fortalecimiento Institucional;
CONACyT M\'exico under Grants No.~232893 (sabbatical), No. 167335, and
No. 179881; Fundaci\'on Marcos Moshinsky; and the Instituto Avanzado
de Cosmolog\'ia Collaboration.
\end{acknowledgments}

\appendix
\section{Notes on the accuracy of the series expansion in
  Eq.~\eqref{eq:5} \label{sec:notes-accur-seri}}
A shown in the previous sections, the inflationary solutions of the
equations of motion~\eqref{eq:4}, under the ansatz~\eqref{eq:5} for
the potential variabls $y_I$, give accurate enough results of
the inflationary variables if the series expansion of the first
potential variable $y_1$ is considered up to the third order. This is
also the highest order at which one can find analytical expressions of
Eq.~\eqref{eq:6}, namely Eqs.~\eqref{eq:27}.

It is certainly expected that the solutions will improve if more
higher-order terms are taken into account, but the question remains of
whether there is any \emph{a priori} assessment of the accuracy of the
method, given that at the end of inflation $\theta_{end} =
\mathcal{O}(1)$. The following is a brief discussion about this issue.

To begin with, the truncated series expansion $y_1 \simeq k_{11}
\theta + k_{12} \theta^2 + k_{13} \theta^3$ indicates that at the end
of inflation $y_1(\theta_{end}) \sim 3 \theta_{end} \simeq 3.69$, with
the precise value depending upon the exact form of the potential. The
problem is that we do not know beforehand what is the true value of
$y_1(\theta_{end})$, and then we do not have a value of reference to
compare our results with.

The most we can do is to estimate $y_1(\theta_{end})$ by other means,
like in the case of the slow-roll approximation (SRA). As is known,
under the SRA the end of inflation is estimated to happen by the
breaking of the slow-roll condition; the latter is given by
\begin{equation}
  \label{eq:23}
  \epsilon_V = \frac{1}{2\kappa^2} \left( \frac{\partial_\phi V}{V}
  \right)^2 = \frac{y^2_1}{12 y^2} \simeq 1 \, ,
\end{equation}
where $\epsilon_V$ is the so-called first slow-roll parameter. If we
assume that Eq.~\eqref{eq:23} really coincides with the end of
inflation at $\theta_{end}$, we find that $y_1(\theta_{end}) \simeq
2\sqrt{3} \cos(\theta_{end}/2)$. We must notice that the breaking of
the slow-roll condition in Eq.~\eqref{eq:23} is the same for all
potentials, which means that under the SRA we must in general expect
that $y_1(\theta_{end}) \simeq 1.15$.

As rough as it is, the slow-roll expression~\eqref{eq:23} seems to
give us a correct order-of-magnitude estimation of the value of $y_1$
at the end of inflation. For instance, the end value provided by the
aforementioned (truncated) expansion of $y_1$ in the case of LFI with
$n=1$ is $y_1(\theta_{end}) = 4.16$ (see
Sec.~\ref{sec:case-ii:-k_11}), whereas in the case of
Starobinsky's model we get $y_1(\theta_{end}) = 5.05$ (see Sec.~\ref{sec:case-iv:-k_11}).

Another hint about the accuracy of the third-order solution comes
from the behavior of the series coefficients in Eq.~\eqref{eq:5}. If
we rewrite the latter in the case of $y_1$ as
\begin{equation}
  \label{eq:31}
  y_1 (\theta) = k_{11} \theta \left( 1 + \frac{k_{12}}{k_{11}} \theta
  + \frac{k_{13}}{k_{11}} \theta^2 + \ldots \right) \, ,
\end{equation}
we can see that the possible convergence of the series~\eqref{eq:31}
depends upon the relative values of the coefficients of higher-order
with respect to the first one $k_{11}$. For the same selected examples
studied in the main text, we find that
\begin{subequations}
  \label{eq:32}
\begin{eqnarray}
  \mathrm{NI}: && \quad \frac{k_{12}}{k_{11}} = 0 \, , \;
  \frac{k_{13}}{k_{11}} = \frac{1}{8} \, , \label{eq:32a} \\
  \mathrm{LFI} (n=1): && \quad \frac{k_{12}}{k_{11}} = 0 \, , \;
  \frac{k_{13}}{k_{11}} = \frac{1}{12} \, , \label{eq:32b} \\
  \mathrm{Starobinsky}: && \quad \frac{k_{12}}{k_{11}} = \frac{1}{3} \, , \;
  \frac{k_{13}}{k_{11}} = - \frac{1}{36} \, . \label{eq:32c}
\end{eqnarray}
\end{subequations}
In the case of the value at the end of inflation, $y_1(\theta_{end})$,
Eqs.~\eqref{eq:32} seem to indicate that the terms inside the brackets
have a decreasing contribution to the total sum. Actually, the third-order term $(k_{13}/k_{11})\theta^2_{end}$ in Eq.~\eqref{eq:31}
contributes the most in the case of NI, but only with a correction of
about $18$\%, and then it is reasonable to think that higher-order
terms will have a less important contribution than that.

The same conclusion seems to arise from the general case in
Eqs.~\eqref{eq:14}. For the purposes of illustration, we consider that
$k_{11}=3$ ($k_{20}=0$), and then
\begin{equation}
  \label{eq:34}
  \left. \frac{k_{12}}{k_{11}} \right|_{k_{11}=3} = \frac{k_{21}}{18} \, , \quad
  \left. \frac{k_{13}}{k_{11}} \right|_{k_{11}=3} = \frac{1}{12} -
  \frac{k^2_{21}}{54} + \frac{k_{22}}{6} \, .
\end{equation}
Although not a general proof because we lack the information about
the coefficients $k_{2j}$ [see Eqs.~\eqref{eq:26}], Eqs.~\eqref{eq:34}
reinforce the idea that an expansion of $y_1$ up to the third order in
Eq.~\eqref{eq:5} may well suffice to obtain accurate enough
inflationary solutions.

\bibliography{axiverse-refs}

\end{document}